\documentstyle[12pt]{article}
\def\II{{\rm 1\!\!I}}
\def\RR{{\rm I\!R}}
\def\NN{{\bf N}}

\def\tr{{\rm tr\,}}
\def\Tr{{\rm Tr\,}}

\def\det{{\rm det\,}}

\def\Det{{\rm Det\,}}

\def\sinh{{\rm sinh\,}}

\def\coth{{\rm coth\,}}

\def\log{{\rm log\,}}

\def\vol{{\rm vol\,}}
\def\R{{\cal R}}

\def\End{{\rm End\,}}

\def\a{\alpha}
\def\b{\beta}
\def\g{\gamma}
\def\d{\delta}

\def\l{\lambda}
\def\m{\mu}
\def\n{\nu}

\def\om{\omega}
\def\na{{\nabla}}

\def\h#1{{\cal #1}}
\def\mathbb#1{{\bf #1}}
\def\boxtimes{{\sq\!\!\!\!\!\!\times}}

\def\sqr#1{\mathop{\mkern0.5\thinmuskip
\vbox{\hrule\hbox{\vrule
\hskip#1\vrule height#1 width 0pt\vrule}\hrule}
\mkern0.5\thinmuskip}}
\def\Square{\mathchoice{\sqr{6pt}}{\sqr{6pt}}
{\sqr{4pt}}{\sqr{3pt}}}
\def\sq{\Square}

\def\be{\begin{equation}}
\def\ee{\end{equation}}
\def\bea{\begin{eqnarray}}
\def\eea{\end{eqnarray}}

\begin{document}

\begin{titlepage}
\null\vskip-3truecm\hspace*{7truecm}{\hrulefill}\par\vskip-4truemm\par
\hspace*{7truecm}{\hrulefill}\par\vskip5mm\par
\hspace*{7truecm}{{\large\sf University of Greifswald (March, 1997)}}
\vskip4mm\par\hspace*{7truecm}{\hrulefill}\par\vskip-4truemm\par
\hspace*{7truecm}{\hrulefill}\par
\bigskip\bigskip
\hspace*{7truecm} 
hep-th/9704166
\bigskip
\bigskip
\bigskip\par
\hspace*{7truecm}
Extended version of a lecture\par
\hspace*{7truecm}
presented at the University of Iowa,
\par
\hspace*{7truecm}
Iowa City, April 4, 1997
\bigskip
\par
\vspace{3cm}
\centerline{\LARGE\bf Covariant techniques}
\centerline{\LARGE\bf for computation of the heat kernel}
\bigskip
\centerline{\Large\bf I. G. Avramidi
\footnote{On leave of absence from Research Institute for Physics, 
Rostov State University,  Stachki 194, 344104 Rostov-on-Don, Russia.}
\footnote{\sc E-mail: avramidi@rz.uni-greifswald.de}}
\bigskip
\centerline{\it Department of Mathematics, University of Greifswald}
\centerline{\it
F.-L.-Jahnstr. 15a, D--17489 Greifswald, Germany}
\bigskip\centerline{\today}\medskip
\bigskip\bigskip
{\narrower\par

The heat kernel associated with an elliptic second-order partial 
differential operator of Laplace type acting on \hbox{smooth} 
sections of a vector bundle over a Riemannian manifold, is studied.
A general manifestly covariant method for computation of the
coefficients of the heat kernel asymptotic expansion is developed.
The technique enables one to compute explicitly the diagonal values 
of the heat kernel coefficients, so called 
Hadamard-Minackshisundaram-De Witt-Seeley coefficients, 
as well as their derivatives.
The elaborated technique is applicable for a manifold of arbitrary dimension
and for a generic Riemannian metric of arbitrary signature.
It is very algorithmic, and well suited to automated computation.
The fourth heat kernel coefficient is computed explicitly 
for the first time.

The general structure of the heat kernel coefficients is investigated 
in detail. On the one hand, the leading derivative terms in all 
heat kernel coefficients are computed. 
On the other hand, the generating functions in closed 
covariant form for the covariantly constant terms and some 
low-derivative terms in the heat kernel coefficients are constructed
by means of purely algebraic methods. 
This gives, in particular, the whole sequence of heat kernel coefficients for an 
arbitrary locally symmetric space.

\bigskip
\par}
\vfill
\end{titlepage}

\section{Introduction}

In this talk I am going to report on recent progress on developing some
computational methods for the heat kernel that turned out to be very powerful
for carrying out explicit computations
\cite{avr86b,avr91b,avr93b,avr95b,avr95c,avr94a,avr96a}.
We will start with the definition of the heat kernel and, then, will try to explain the main
ideas of our approach and present the main results without going much into details.
Consequently, I will not be as formal as probably many matematicians would like 
me to be.

The heat kernel proved to be a very powerful tool in mathematical physics
as well as in quantum field theory. It has been the subject of much investigation
in recent years in mathematical as well as in physical literature.
(See, for example, \cite{gilkey95,berline92,tam95,avr86b,avr91b,avr95c} 
and references therein.) 
The study of the heat kernel is motivated, in particular, by the fact that it gives 
a general framework of
covariant methods for investigating the quantum field theories with local gauge
symmetries, such as quantum gravity and gauge theories
\cite{barvinsky85}.

\subsection{Preliminaries}

To define the heat kernel one has to remember some preliminary facts from
the differential geometry \cite{berline92}.
Let $(M,g)$ be a smooth Riemannian manifold of dimension $d$ with a positive
definite Riemannian metric $g$.
To simplify the
exposition we assume additionally that it is {\it compact} and 
{\it complete}, i.e. 
without boundary, $\partial M=\emptyset$.

Let $TM$ and $T^*M$ be the tangent and cotangent bundles of the manifold $M$.
On the tangent bundle $TM$ of a Riemannian manifold there is always a
unique canonical connection, so called {\it Levi-Civita connection}, 
$\nabla^{TM}$, which is torsion-free and compatible with the metric $g$.
 
Let $V$ be a smooth vector bundle over the manifold $M$,
${\rm End}\,(V)$ be the bundle of all smooth
endomorphisms of the vector bundle $V$, 
and
$C^\infty(M,V)$ and $C^\infty(M,{\rm End}\,(V))$ be the spaces of all 
smooth sections of the vector bundles $V$ and ${\rm End}\,(V)$. 

Further, we will also assume that $V$ is a {\it Hermitian} vector bundle, i.e. 
there is a Hermitian pointwise fibre scalar product $<\varphi, \psi >$ for any two
sections of the vector bundle $\varphi, \psi \in C^\infty(M,V)$.
The dual vector bundle $V^*$ is naturally identified with $V$, so that
\be
<\varphi,\psi>=\tr_V(\bar\varphi\otimes\psi),
\ee
where $\psi\in C^\infty(M,V)$, and $\bar\varphi\in C^\infty(M,V^*)$ and $\tr_V$ is
the fibre trace.
Using  the invariant Riemannian volume element $d\vol(x)$ on the manifold $M$
we define a natural $L^2$ {\it inner product}
\be
(\varphi,\psi)=\Tr_{L^2}(\bar\varphi\otimes\psi)
=\int\limits_M d\vol(x) <\varphi,\psi>
=\int\limits_M d\vol(x) \tr_V(\bar\varphi\otimes\psi).
\ee
The Hilbert space $L^2(M,V)$ is defined to be the completion of $C^\infty(M,V)$ in this
norm. 

Let $\nabla^V$ be a connection, or covariant derivative, 
on the vector bundle $V$
\be
\nabla^V:\ C^\infty(M,V)\to C^\infty(M,T^*M\otimes V),
\ee
which is {\it compatible with the Hermitian metric} on the vector bundle $V$, i.e.
\be
\nabla<\varphi,\psi>=<\nabla^V\varphi,\psi>+<\varphi,\nabla^V\psi>.
\label{100}
\ee

On the tensor product bundle $T^*M\otimes V$ we define
the tensor product connection by means of the Levi-Civita connection
\be
\nabla^{T^*M\otimes V}=\nabla^{T^*M}\otimes 1+1\otimes\nabla^{V}.
\ee
Similarly, we extend the connection $\nabla^V$ with the help of the Levi-Civita connection
to $C^\infty(M,V)$-valued tensors of all orders and denote it just by $\nabla$. 
Usually there is no
ambiguity and the precise meaning of the covariant derivative is always clear from
the nature of the object it is acting on.

The composition of two covariant derivatives is a mapping
\be
\nabla^{T^*M\otimes V}\nabla^V:\ 
C^\infty(M,V)\to C^\infty(M,T^*M\otimes V)\to
C^\infty(M,T^*M\otimes T^*M\otimes V).
\ee
Let, further, ${\rm tr}_g$ denote the contraction of sections of the bundle 
$T^*M\otimes T^*M\otimes V$ with the metric on the cotangent bundle
\be
{\rm tr}_g\,=g\otimes 1 :\ C^\infty(M,T^*M\otimes T^*M\otimes V)\to
C^\infty(M,V).
\ee
Then we can define a second-order differential operator, called 
the {\it generalized Laplacian}, by
\be
\sq={\rm tr}_g \nabla^{T^*M\otimes V}\nabla^V
\ee
\be
\sq\,:\
C^\infty(M,V)\to C^\infty(M,T^*M\otimes V)\to
C^\infty(M,T^*M\otimes T^*M\otimes V)\to
C^\infty(M,V).
\ee
Further, let $Q$ be a a smooth {\it Hermitian} section of the endomorphism bundle, 
${\rm End}\,(V)$, i.e.
\be
<\varphi,Q\psi>=<Q\varphi,\psi>.
\ee
Finally,
we define a {\it Laplace type differential operator} $F$ as the sum of the 
generalized Laplacian and the endomorphism $Q$
\be
F=-\sq+Q.
\ee

\subsection{Laplace type operator in local coordinates}

The generalized Laplacian can be easily expressed in local coordinates. 
Let $x^\mu$, $(\mu=1,2,\dots, d)$, be a system of local coordinates and
$\partial_\mu$ and $dx^\mu$ be the local coordinate frames for 
the tangent and the cotangent bundles.
We adopt the notation that the Greek indices label the 
tensor components with respect to local 
coordinate frame and range from 1 through $d=\dim M$. Besides, 
a summation is always caried out over repeated indices.
Let $g_{\mu\nu}=(\partial_\mu,\partial_\nu)$ be the metric on the tangent bundle, 
$g^{\mu\nu}=(dx^\mu,dx^\nu)$ be the metric on the cotangent bundle, 
$g=\det g_{\mu\nu}$,
$\Gamma^\mu_{\nu\lambda}$ be the Levi-Civita connection
and ${\cal A}_\mu$ be the connection $1-$form of $\nabla^V$.

Then it is not difficult to obtain for the generalized Laplacian 
\be
\sq=g^{\mu\nu}\nabla_\mu\nabla_\nu=
g^{-1/2}(\partial_\mu+{\cal A}_\mu)g^{1/2}g^{\mu\nu}
(\partial_\nu+{\cal A}_\nu).
\ee
Therefore, a Laplace type operator is a second-order partial differential operator
of the form
\be
F=-g^{\mu\nu}\partial_\mu\partial_\nu-2a^\mu\partial_\mu+q,
\ee
where $a^\mu$ is a $\End(V)$-valued vector
\be
a^\mu=g^{\mu\nu}{\cal A}_\nu
+{1\over 2}g^{-1/2}\partial_\nu(g^{1/2}g^{\nu\mu})
\ee
and $q$ is a section of the endomorphism bundle $\End(V)$
\be
q=Q-g^{\mu\nu}{\cal A}_\mu{\cal A}_\nu
-g^{-1/2}\partial_\mu(g^{1/2}g^{\mu\nu}{\cal A}_\nu).
\ee

Thus, a Laplace type operator is constructed from the following three pieces
of geometric data
\begin{itemize}
\item
a {\it metric} $g$ on $M$, which determines the second-order part;
\item
a {\it connection} $1$-form ${\cal A}$ on the vector bundle $V$, 
which determines the first-order
part;
\item
an {\it endomorphism} $Q$ of the vector bundle $V$, which determines the zeroth order
part.
\end{itemize}
It is worth noting that every second-order differential operator with a scalar 
leading symbol given by the metric tensor is of Laplace type and 
can be put in this form by choosing the appropriate connection $\nabla^V$ 
and the endomorphism $Q$.

\subsection{Self-adjoint operators}

Using the $L^2$ inner product we define the {\it adjoint} $F^*$ of a 
differential operator $F$ by
\be
(F^*\varphi,\psi)=(\varphi,F\psi).
\ee

It is not difficult to prove that if the connection $\nabla$ is compatible with the
Hermitian metric on the vector bundle $V$ and the boundary of the manifold $M$
is empty, then the generalized Laplacian $\sq$, and, obviously, 
any Laplace type operator $F$,
is an {\it elliptic symmetric} differential operator
\be
(\sq\varphi,\psi)=(\varphi,\sq\psi), 
\qquad 
(F\varphi,\psi)=(\varphi,F\psi),
\ee
with a positive principal symbol.
Moreover, the operator $F$ is {\it essentially self-adjoint}, i.e. there is a unique 
{\it self-adjoint extension} $\bar F$ of the operator $F$.
We will not be very careful about distinguishing 
between the operator $F$ and its closure $\bar F$, and will simply say that the operator
$F$ is elliptic and self-adjoint.

\paragraph{Spectral theorem.}

There is a well known theorem about the spectrum of 
any elliptic self-adjoint differential operator $F$ acting on smooth sections of a vector
bundle $V$ over a compact manifold $M$,
$F: C^\infty(M,V)\to C^\infty(M,V)$, with a positive definite principal symbol
\cite{gilkey95}.
Namely,
\begin{itemize}
\item
the operator $F$ has a discrete real spectrum, $\lambda_n$, ($n=1,2,\dots,$), 
bounded from below
\be
\lambda_n>-C, 
\ee
with some real sonstant $C$,
\item
all eigenspaces of the operator $F$ are finite-dimensional and
the eigenvectors, $\varphi_n$,  of the operator $F$, 
\be
F\varphi_n=\lambda_n\varphi_n,
\ee
are smooth sections of the vector bundle $V$,
which form a complete orthonormal basis in $L^2(M,V)$.
\be
(\varphi_n,\varphi_m)=\delta_{mn}.
\ee
\end{itemize}

In the following we wil assume that  the endomorphism $Q$ is bounded from below by
a sufficiently large constant, so that the Laplace type operator $F$ is {\it strictly positive}. 
This does not
influence all the conclusions but simplifies significantly the technical details needed
to treat the negative and zero modes of the operator $F$. This can be always done
as long as we study only asymptotic properties of the spectrum for large eigenvalues but
not the structure and the dimension of the null space and related cohomolgical and
topological questions.

\subsection{Heat kernel}

Thus all eigenfunctions of the Laplace type operator $F$ are smooth 
sections of the vector bundle $V$ and, if the manifold $M$ is compact, 
$F$ has a unique self-adjoint extension, which 
we denote by the same symbol $F$. 

Then the operator $U(t)=\exp(-tF)$ for $t>0$ is well defined as
a {\it bounded} operator on the Hilbert space of square integrable sections of the vector bundle
$V$. These operators form a one-parameter semi-group.
The kernel $U(t|x,x')$ of this operator is defined by
\be
U(t|x,x')=\exp(-tF)\delta(x,x')
=\sum\limits_n e^{-t\lambda_n}\varphi_n(x)\otimes \bar\varphi_n(x'),
\ee
where $\delta(x,x')$ is the covariant Dirac distribution along the diagonal of $M\times M$,
and each eigenvalue is counted with multiplicities. 
It can be regarded as an endomorphism from the fiber of $V$ over $x'$ to the
fiber of $V$ over $x$.

The kernel $U(t|x,x')$ of the operator $\exp(-tF)$ satisfies the heat equation
\be
(\partial_t+F)U(t|x,x')=0
\ee
with the initial condition
\be
U(0^+|x,x')=\delta(x,x').
\ee
That is why, it is called the {\it heat kernel}.
It can be proved that there is a unique smooth solution, called  the {\it fundamental solution}, 
of the heat equation satisfying that initial condition. 
Thus, the heat kernel is the {\it fundamental solution} of the heat equation.
{}For $t>0$ the heat kernel is a smooth section of the external tensor product
of the vector bundles $V\boxtimes V^*$ over the tensor product manifold $M\times M$:\
$U(t|x,x')\in C^\infty(\RR_+\times M\times M, V\boxtimes V^*)$.

It is not difficult to prove that for a positive elliptic operator $F$ the inverse operator,
called also Green operator,
$G=F^{-1}$, is a bounded operator with the kernel given by
\be
G(x,x')=\int\limits_0^\infty dt\,U(t|x,x').
\ee

\subsection{Trace of the heat kernel and the spectral functions}

As we already said above, for all $t>0$ the heat semi-group $U(t)=\exp(-tF)$ of a Laplace type
operator $F$ on a compact manifold $M$ is a bounded operator on the Hilbert
space $L^2(M,V)$ and is {\it trace-class}, 
with a well defined {\it trace} given by the formula
\be
\Tr_{L^2}\exp(-tF)=\int\limits_M d\vol(x)\tr_V U(t|x,x)
=\sum\limits_n e^{-t\lambda_n}.
\ee

The trace of the heat kernel is obviously a spectral invariant of the operator $F$.
It determines all other spectral functions by integral transforms.

\begin{enumerate}
\item
The {\it distribution function}, $N(\lambda)$, defined as the number of the eigenvectors 
with the eigenvalues less than $\lambda$, is given by
\be
N(\lambda)=\#\{\varphi_n|\ \lambda_n<\lambda\}
={1\over 2\pi i}\int\limits_{c-i\infty}^{c+i\infty}{dt\over t}
e^{t\lambda}\,\Tr_{L^2}\exp(-tF),
\ee
where $c$ is a positive constant.
\item
The {\it density function}, $\rho(\lambda)$, 
is defined by derivative of the distribution function and is
obviously
\be
\rho(\lambda)={d\over d\lambda}N(\lambda)
={1\over 2\pi i}\int\limits_{c-i\infty}^{c+i\infty}dt
e^{t\lambda}\,\Tr_{L^2}\exp(-tF).
\ee
\item
The zeta-function, $\zeta(s)$, defined as the trace of the complex power of the operator $F$,
is given by
\be
\zeta(s)=\Tr_{L^2}F^{-s}=\int d\lambda\rho(\lambda)\lambda^{-s}
={1\over\Gamma(s)}\int\limits_0^\infty dt t^{s-1}\,\Tr_{L^2}\exp(-tF),
\ee
where $s$ is a complex variable with ${\rm Re}\, s>d/2$.
\item
Finally, the trace of the powers of the resolvent operator is also expressed in terms
of the heat kernel
\be
R_k(z)=\Tr_{L^2}(F-z)^{-k}=\int d\lambda\rho(\lambda)(\lambda-z)^{-k}
={1\over (k-1)!}\int\limits_0^\infty dt t^{k-1}e^{tz}\,\Tr_{L^2}\exp(-tF),
\ee
where $z$ is a complex parameter with ${\rm Re}\, z<0$ and 
$k$ is a positive integer satisfying the condition $k>d/2$.
\end{enumerate}

These spectral functions are very useful tools in studying the spectrum of the operator $F$.
The zeta function enables one to define, in particular, the regularized determinant
of the operator $F$, 
\be
\zeta'(0)=-\log\Det F,
\ee
which determines the one-loop effective action in quantum field theory.
All these functions are, in principle, equivalent to each other. However, the heat
kernel is a smooth function whereas the distribution and especially 
the density function are extremely singular. That is why the heat kernel seems to be
more convenient for practical purposes.

\section{Asymptotic expansion of the heat kernel}

In the following we are going to study the heat kernel only locally, i.e.
in the neighbourhood of the diagonal of $M\times M$, when the points $x$ and $x'$ are
close to each other. The exposition will follow mainly
our papers \cite{avr86b,avr91b}.
We will keep a point $x'$ of the manifold fixed and consider a
small geodesic ball, i.e. a small neighbourhood of the point $x'$: 
$B_{x'}=\{x\in M| r(x,x')<\varepsilon\}$, $r(x,x')$ being the geodesic distance between the 
points $x$ and $x'$. We will take the radius of the ball sufficiently small, so that each 
point $x$ of the ball
of this neighbourhood can be connected by a unique geodesic 
with the point $x'$. This can be always
done if the size of the ball is smaller than the injectivity radius of the manifold at $x'$,
$\varepsilon<r_{\rm inj}(x')$.

Let $\sigma(x,x')$ be the geodetic interval, also caled {\it world function}, 
defined as one half the square of the length
of the geodesic connecting the points $x$ and $x'$
\be
\sigma(x,x')={1\over 2}r^2(x,x').
\ee
The first derivatives of this function with respect to $x$ and $x'$ 
define tangent vector fields to the geodesic at the points
$x$ and $x'$
\be
u^\mu=g^{\mu\nu}\nabla_\nu\sigma,\qquad
u^{\mu'}=g^{\mu'\nu'}\nabla'_{\nu'}\sigma.
\ee 
and the determinant of the mixed second derivatives defines a so called 
{\it Van Vleck-Morette determinant}
\be
\Delta(x,x')=g^{-1/2}(x)\det(-\nabla_\mu\nabla'_{\nu'}\sigma(x,x'))g^{-1/2}(x').
\ee
Let, finally, ${\cal P}(x,x')$ denote the paralell transport operator along the geodesic
from the point $x'$ to the point $x$. It is a section of the external tensor product of the
vector bundle $V\boxtimes V^*$ over $M\times M$, or, in other words, it is an
endomorphism from the fiber of $V$ over $x'$ to the fiber of $V$ over $x$.

Near the diagonal of $M\times M$ all these two-point functions are smooth single-valued 
functions of the coordinates of the points $x$ and $x'$. To simplify the consideration, 
we will assume that these functions are analytic.

Then, the function
\be
U_0(t|x,x')=(4\pi t)^{-d/2}\Delta(x,x')\exp\left(-{1\over 2t}\sigma(x,x')\right){\cal P}(x,x')
\ee
satisfies the initial condition
\be
U_0(0^+|x,x')=\delta(x,x').
\ee
Moreover, {\it locally} it satisfies also 
the heat equation in the {\it free} case, when the Riemannian 
curvature of the manifold ${\rm Riem}$, 
the curvature of the bundle connection ${\cal R}$ and the endomorphism
$Q$ vanish: ${\rm Riem}={\cal R}=Q=0$. 
Therefore, $U_0(t|x,x')$ is the exact heat kernel for a pure generalized Laplacian
in flat Euclidean space with a flat
trivial bundle connection and without the endomorphism $Q$.

This function gives a good framework for the approximate solution in the general case.
Namely, by factorizing out this free factor we get an ansatz 
\be
U(t|x,x')=(4\pi t)^{-d/2}\Delta(x,x')\exp\left(-{1\over 2t}\sigma(x,x')\right){\cal P}(x,x')
\Omega(t|x,x').
\label{150}
\ee
The function $\Omega(t|x,x')$, called the {\it transport function}, is a section of the
endomorphism vector bundle $\End(V)$ over the point $x'$.
Using the definition of the functions $\sigma(x,x')$, $\Delta(x,x')$ and ${\cal P}(x,x')$
it is not difficult to find that the transport function satisfies a transport equation
\be
\left(\partial_t+{1\over t}D+L\right)\Omega(t)=0,
\ee
where $D$ is the radial vector field, i.e. operator of differentiation along the geodesic,
defined by
\be
D=\nabla_u=u^\mu\nabla_\mu,
\ee
and $L$ is a second-order differential operator defined by
\be
L={\cal P}^{-1}\Delta^{-1/2}F\Delta^{1/2}{\cal P}.
\label{160}
\ee
The initial condition for the transpot function is obviously
\be
\Omega(t|x,x')=I,
\ee
where $I$ is the identity endomorphism of the vector bundle $V$ over $x'$.

One can prove that when the operator $F$ is positive the function $\Omega(t)$
satisfies the following asymptotic conditions
\be
\lim_{t\to\infty,0}t^\alpha\partial_t^N\Omega(t)=0 \qquad
{\rm for\ any}\ \alpha>0,\ N\ge 0.
\label{200}
\ee
In other words, as $t\to\infty$ the function $\Omega(t)$ and all its derivatives
decreases faster than any power of $t$, actually it decreases exponentialy, and 
as $t\to 0$ the product of $\Omega(t)$ with any positive power of $t$ vanishes.

Now, let us consider a slightly modified version of the Mellin transform of the 
function $\Omega(t)$ introduced in \cite{avr91b}
\be
b_q={1\over \Gamma(-q)}\int_0^\infty dt t^{-q-1}\Omega(t).
\ee
This integral converges for ${\rm Re}\, q<0$.
By integrating by parts $N$ times and using the asymptotic conditions (\ref{200})
we get also
\be
b_q={1\over \Gamma(-q+N)}\int_0^\infty dt t^{-q-1+N}(-\partial_t)^N\Omega(t).
\ee
This integral converges for ${\rm Re}\,q<N-1$.
 
Using this representtion one can prove that \cite{avr91b}
\begin{itemize}
\item
the function $b_q$ is analytic everywhere, i.e. it is an entire function,
\item
the values of the function $b_q$ at the integer positive points are given by
\be
b_k=(-\partial_t)^k\Omega(t)\Big|_{t=0},
\ee
\item
$b_q$ satisfies an asymptotic condition
\be
\lim_{|q|\to\infty,\ {\rm Re}\,q<N}\Gamma(-q+N)b_q=0,\qquad
{\rm for\ any}\ N>0.
\label{350}
\ee
\end{itemize}

By inverting the Mellin transform we obtain a new ansatz for the transport function
and, hence, for the heat kernel
\be
\Omega(t)={1\over 2\pi i}\int\limits_{c-i\infty}^{c+i\infty}dq\,t^q\,\Gamma(-q)b_q
\ee
where $c<0$ is a negative constant. 

Substituting this ansatz into the transport equation we get a functional eqution for 
the function $b_q$
\be
\left(1+{1\over q}D\right)b_q=L\,b_{q-1}.
\label{400}
\ee
The initial condition for the transport function is translated into
\be
b_0=I.
\label{500}
\ee

Thus, we have reduced the problem of solving the heat equation to the following problem:
one has to find an entire function $b_q(x,x')$ that satisfies the functional equation
(\ref{400}) with the initial condition (\ref{500}) and the asymptotic condition (\ref{350}).

The function $b_q$ turns out to be extremely useful in computing the heat kernel, the resolvent
kernel, the zeta-function and the determinant of the operator $F$. It contains the same 
the information about the manifold as the heat kernel.
In some cases the function $b_q$ can be constructed just by analytical continuation from
the integer positive values $b_k$ \cite{avr91b}.

Now we are going to do the usual trick, namely, to move the contour of integration over $q$
to the right. Due to the presence of the gamma function $\Gamma(-q)$ the integrand 
has simple poles at the non-negative integer
points $q=0,1,2\dots$, which contribute to the integral while moving the contour.
So, we get
\be
\Omega(t)=\sum\limits_{k=0}^{N-1}{(-t)^k\over k!}b_k+R_N(t),
\label{300}
\ee
where
\be
R_N(t)={1\over 2\pi i}\int\limits_{c_N-i\infty}^{c_N+i\infty}dq\,t^q\,\Gamma(-q)b_q
\ee
with $c_N$ is a constant satisfying the condition $N-1<c_N<N$. 
As $t\to 0$ the rest term $R_N(t)$ behaves like $O(t^N)$, so we obtain an
asymptotic expansion as $t\to 0$
\be
\Omega(t|x,x')\sim \sum\limits_{k\ge 0}{(-t)^k\over k!}b_k(x,x').
\label{600}
\ee

Using our ansatz (\ref{150}) we find immediately the trace of the heat kernel
\be
\Tr_{L^2}\exp(-tF)=(4\pi t)^{-d/2}
{1\over 2\pi i}\int\limits_{c-i\infty}^{c+i\infty}dq\,t^q\,\Gamma(-q)B_q,
\ee
where
\be
B_q=\Tr_{L^2}b_q=\int\limits_M d\vol(x)\tr_V\,b_q(x,x).
\ee
The trace of the heat kernel has an analogous asymptotic expansion as $t\to 0$
\be
\Tr_{L^2}\exp(-tF)\sim (4\pi t)^{-d/2}\sum\limits_{k\ge 0}{(-t)^k\over k!}B_k.
\ee

This is the famous Minackshisundaram-Pleijel asymptotic expansion.
The physicists call it the Schwinger-De Witt expansion \cite{barvinsky85}. 
Its coefficients $B_k$ are also called sometimes 
Hadamard-Minackshisundaram-De Witt-Seeley (HMDS) coefficients.
This expansion is of great importance in differential geometry, spectral geometry, 
quantum field theory and other areas of mathematical physics, such as 
theory of Huygence' principle, heat kernel proofs of the index theorems, 
Korteveg-De Vries hierarchy, Brownian motion etc. (see, forexample, 
\cite{hurt83}).

{}For integer $q=k=1,2,\dots$ the functional equation (\ref{400}) becomes a recursion system
that, together with the initial condition (\ref{500}),
determines all the HMDS-coefficients $b_k$.

One should stress, however, that this series does not converge, in general. 
In that sense our ansatz (\ref{300}) in form of a Mellin transform of an entire 
function is much better since it is exact and gives an explicit formula for the rest term.

Let us apply our ansatz for computation of the complex power of the operator $F$
defined by
\be
G^p=F^{-p}={1\over\Gamma(p)}\int\limits_0^\infty dt\, t^{p-1}\,U(t).
\ee
Using our ansatz for the heat kernel we obtain
\be
G^p=(4\pi)^{-d/2}\Delta^{1/2}{\cal P}{1\over 2\pi i}\int\limits_{c-i\infty}^{c+i\infty}
dq\,{\Gamma(-q)\Gamma(-q-p+d/2)\over\Gamma(p)}
\left(\sigma\over 2\right)^{q+p-d/2}b_q
\ee
where $c<-{\rm Re}\,p+d/2$.

Outside the digonal, i.e. for $\sigma\ne 0$, this integral converges for any $p$ 
and defines an entire function of $p$.
The integrand in this formula is a meromorphic function of $p$ with some simple and
maybe some double poles. If we move the contour of integrtion to the right, we get
contributions from the simple poles in form of powers of $\sigma$
and a logarithmic part due to the double poles (if any).
This gives the complete structure of diagonal singularities of the complex power of the opertor 
$F$, $G^p(x,x')$.
Thus the function $b_q$ turns out to be very useful to study the diagonal singularities.

In particular case $p=1$ we recover in this way the singularity structure of the Green function
\be
G=(4\pi)^{-d/2}\Delta^{1/2}{\cal P}\left(\Phi+\Psi\,\log{\sigma\over 2}\right)
+G_{\rm reg},
\ee
where
\be
\Phi=\sum\limits_{k=0}^{d/2-1}{(-1)^k\over k!}\Gamma(d/2-1-k)
\left(2\over\sigma\right)^{d/2-1-k}b_k
\ee
\be
\Psi=\left\{
\begin{array}{ll}
0, & {\rm for \ odd}\ d\\
{\displaystyle {(-1)^{d/2}\over\Gamma(d/2)}}b_{d/2-1} & {\rm for\ even\ } d
\end{array}
\right.
\ee
\be
G_{\rm reg}=(4\pi)^{-d/2}\Delta^{1/2}{\cal P}{1\over 2\pi i}
\int\limits_{\alpha-i\infty}^{\alpha+i\infty}dq\,\Gamma(-q)\Gamma(-q-1+d/2)
\left(\sigma\over 2\right)^{q+1-d/2}b_q,
\ee
where $[d/2]-1<\alpha<[d/2]-1/2$.
We see that due to the absence of the double poles in the integrand 
there is no logarithmic singularity in odd dimensions.
Thus, all the singularities of the Green function 
are determined by the HMDS-coefficients $b_k$ and the regular part is determined by
the function $b_q$ as well.

Now, let us consider the diagional limit of $G^p$. By taking the limit $\sigma\to 0$ we 
obtain a very simple formula in terms of the function $b_q$
\be
G^p(x,x)=(4\pi)^{-d/2}{\Gamma(p-d/2)\over\Gamma(p)}b_{d/2-p}(x,x).
\ee
This gives automatically the zeta-function of the operator $F$ \cite{avr91b}
\be
\zeta(p)=(4\pi)^{-d/2}{\Gamma(p-d/2)\over\Gamma(p)}B_{d/2-p}.
\ee

Herefrom we see that both $G^p(x,x)$ and $\zeta(p)$ are meromorphic functions 
with simple poles at the points $p=[d/2]+1/2-k$, $(k=0,1,2,\dots)$ and $p=1,2,\dots, [d/2]$.
In particular, the zeta-function is analytic at the origin. Its value at the origin is given
by
\be
\zeta(0)=\left\{
\begin{array}{ll}
0 & {\rm for\ odd\ } d\\
(4\pi)^{-d/2}{\displaystyle {(-1)^{d/2}\over\Gamma(d/2+1)}}
B_{d/2} & {\rm for \ even \ } d
\end{array}
\right.
\ee
This gives the regularized number of all modes of the operator $F$ 
since formally 
\be
\zeta(0)=\Tr_{L^2} \II=\sum\limits_n 1.
\ee 

Moreover, the derivative of the zeta-function at the origin is also well defined.
As we already mentioned above it determines the regularized determinant of the 
operator $F$ since formally
\be
\log\Det F=\Tr_{L^2}\log F=\sum\limits_n \log\lambda_n=-\zeta'(0).
\ee
Thus we obtain for the determinant
\be
\log\Det F=-(4\pi)^{-d/2}{\pi(-1)^{(d+1)/2}\over\Gamma(d/2+1)}B_{d/2}
\qquad
{\rm for\ odd\ } d
\ee
and
\be
\log\Det F=(4\pi)^{-d/2}{(-1)^{d/2}\over\Gamma(d/2+1)}
\left\{B'_{d/2}-[\Psi(d/2+1)+{\bf C}]B_{d/2}
\right\}
\qquad
{\rm for \ even \ } d.
\ee
Here $\Psi(z)=(d/dz)\log\Gamma(z)$ is the psi-function, 
${\bf C}=-\Psi(1)$ is the Euler constant, and
\be
B'_{d/2}={d\over dq}B_q\Bigg|_{q=d/2}.
\ee

\section{Non-recursive solution of the recursion system}

The main problem we are solving is to compute the HMDS-coefficients, not only
the integrated ones $B_k=\int_M d\vol(x)\tr_V b_k(x,x)$, 
which are determined by the diagonal values of $b_k(x,x)$,
but rather the off-diagonal coefficients  $b_k(x,x')$. They are determined by a recursion
system which is obtained simply by restricting the complex variable $q$ in the eq. (\ref{400})
to positive integer values $q=1,2,\dots$. 
This problem was solved in \cite{avr86b,avr91b} where a systhematic technique for 
calculation of $b_k$ was developed.
The formal solution of this recursion system is
\be
b_k=\left(1+{1\over k}D\right)^{-1}L\left(1+{1\over k-1}D\right)^{-1}L\cdots
\left(1+{1\over 1}D\right)^{-1}L\cdot I.
\ee

So, the problem is to give a precise practical meaning to this formal operator solution.
To do this one has, first of all, to define the inverse operator $(1+D/k)^{-1}$. This can be
done by constructing the complete set of eigenvectors of the operator $D$.
However, first we introduce some auxillary notions from the theory of symmetric tensors.

\subsection{Algebra of symmetric tensors}

Let $\omega^a$ and $e_a$ be the basises in the cotangent $T^*M$ and tangent $TM$ bundles,
$S^n(M)$ be the bundle 
of symmetric contravariant tensors of rank $n$,  
$S_n(M)$ be
the bundle of symmetric $n$-forms and 
$S^n_m(M)=S_m(M)\otimes S^n(M)$ be the bundle of symmetric tensors 
of type $(m,n)$ with the basis
\be
s^{a_1\dots a_m}_{b_1\dots b_n}
=\omega^{(a_1}\otimes\cdots\otimes\omega^{a_m)}
\otimes e_{(b_1}\otimes\cdots\otimes e_{b_n)}.
\ee
where the parenthesis mean the symmetrization over all indices included.

In the space $S^n_n$ there is a natural unity symmetric tensor
\be
I_{(n)}=s^{1\dots n}_{1\dots n},
\ee
which is an identical endomorphism of the vector bundles $S^n$ and $S_n$.

We define the following binary operations on symmetric tensors:
\begin{itemize}
\item[a)]
{\it the exterior symmetric tensor product} $\vee$
\be
\vee:\ S^n_m\times S^i_j\to S^{n+i}_{m+j}
\ee
by
\be
A\vee B=A^{( b_1\dots b_n}_{(a_1\dots a_m}
B^{ b_{n+1}\dots b_{n+i})}_{a_{m+1}\dots a_{m+j})}
s^{a_1\dots a_{m+j}}_{b_1\dots b_{n+i}},
\ee

\item[b)] and
an {\it inner product} $\star$
\be
\star:\ S^n_m\times S^i_n\to S^i_{m},
\ee
by:
\be
A\star B
=A^{ c_1\dots c_n}_{a_1\dots a_m}
B^{ b_{1}\dots b_{i}}_{c_1\dots c_n}
s^{a_1\dots a_{m}}_{b_1\dots b_{i}}.
\ee

\end{itemize}

Further, we define also an {\it exterior symmetric covariant derivative} $s\nabla$
on symmetric tensors
\be
s\nabla:\ S^m_n\to S^m_{n+1}
\ee
by
\be
s\nabla A=\nabla_{( a_1}A^{ b_1\dots b_m}_{a_2\dots a_{n+1})}
s^{a_1\dots a_{n+1}}_{b_1\dots b_m}.
\ee

Everything said above remains true if we consider $\End(V)$-valued symmetric 
tensors, i.e. sections of the vector bundle $S^n_m\otimes\End(V)$, for some vector
bundle $V$ over $M$. The product operations include then the usual endomorphism (matrix)
inner product as well.

\subsection{Covariant Taylor basis}

Let us consider the space ${\cal L}(B_{x'})=\{|f>\equiv f(x,x')|\ x\in B_{x'}\}$ 
of smooth analytic two-point functions
in a small neighbourhod $B_x'$ of the diagonal $x=x'$. 
Here we denote the elements of this space by $|f>$.
Let us define a special set of such
functions $|n>\in {\cal L}(B_{x'})$ labeled by a natural number $n\in\NN$ by
\bea
|0>&=&1
\nonumber\\
|n>&=&{(-1)^n\over n!}\vee^n u',
\qquad (n=1,2,\dots),
\eea
where $u'$ is the tangent vector field to the geodesic conecting the points $x$ and $x'$ at the 
point $x'$ given by the first derivative of the geodetic interval $\sigma$
\be
u'=(g'^{ab}\nabla'_{b}\sigma) e'_{a},
\ee
where prime $'$ denotes the objects and operations at the point $x'$.
The functions $|n>$ are two-point geometric objects, which are scalars at the point $x$ and
symmetric contravariant tensors at the point $x'$, more precisely, they are
sections of the vector bundle $S^n$ over the point $x'$.

Let us define also the dual space of linear functionals
\be
{\cal L}^*(B_{x'})=\{<f|:\ {\cal L}(B_{x'})\to {\mathbb C}\},
\ee
with the basis $<n|$ dual to the basis $|n>$. 
The values of the dual basis functionals on the two-point functions are 
sections of the vector bundle of symmetric forms
$S_n$
defined to be the diagonal values of the symmetric exterior covariant derivative
$s\nabla$
\be
<n|f>=[(s\nabla)^nf],
\ee
where the square brackets mean restriction to the diagonal $x=x'$.

The basis $<n|$ is dual to $|m>$ in the sense that
\be
<n|m>=\delta_{mn}I_{(n)}.
\ee

Using this notation the covariant Taylor series for an 
analytic function $|f>$ can be written in the form
\be
|f>=\sum_{n\ge 0}|n>\star <n|f>.
\ee

Now it is almost obvious that our set of functions $|n>$ forms a complete 
basis in ${\cal L}(B_{x'})$
due
to the fact that there does not exist any nontrivial analytic function
which is 'orthogonal' to all of the eigenfunctions $|n>$.
In other words, an analytic function that is equal to zero together with
all  symmetrized derivatives at the point $x=x'$ is, in fact, 
identically equal to zero in $B_{x'}$.

It is easy to show that these functions satisfy the equation
\be
D|n>=n|n>
\ee
and, hence, are the eigenfunctions of the operator $D$ with positive integer 
eigenvalues.
Note, however, that the space of analytical functions ${\cal L}(B_{x'})$ is 
not a Hilbert space with a scalar product $<f|g>$ defined above 
since there are a lot of analytic functions
for which the norm $<f|f>$ diverges.
If we restrict ourselves to polynomial functions of some order then this problem
does not appear. Thus the space of polynomials is a Hilbert space with the inner product
defined above.

\subsection{Covariant Taylor series for HMDS-coefficients $b_k$}

The complete set of eigenfunctions $|n>$ can be employed to present an
arbitrary linear differential operator $L$
in the form
\be
L=\sum\limits_{m,n\ge 0}|m>\star<m|L|n>\star<n|,
\label{700}
\ee
where $<m|L|n>$
are the `matrix elements' of the operator $L$ that 
are just $\End(V)$-valued symmetric tensors, i.e. sections of the vector bundle
$S^n_m(M)\otimes \End(V)$.
We will not study the question of convergency of the expansion
(\ref{700}). It can be regarded just as a formal series. 
When acting on an analytic function, this series is nothing but
the Taylor series and converges in a sufficiently small region $B_{x'}$.

Now it should be clear that the inverse operator $(1+{1\over k}D)^{-1}$ can be defined
by
\be
\left(1+{1\over k}D\right)^{-1}=\sum_{n\ge 0}{k\over k+n}|n>\star<n|.
\ee
Using this representation together with the analogous one for the operator $L$, (\ref{700}) ,
we obtain a covariant Taylor series for the coefficients $b_k$
\be
b_k=\sum_{n\ge 0}|n>\star<n|b_k>
\ee
with the covariant Taylor coefficients $<n|b_k>$ given by \cite{avr86b,avr91b}
\begin{eqnarray}
\lefteqn{<n|b_k>
=\sum_{n_1,\dots,n_{k-1}\ge 0}{k\over k+n}\cdot
{k-1\over k-1+n_{k-1}}\cdots{1\over 1+n_1}}
\qquad\ \,\nonumber\\[10pt]
& &\times <n|L|n_{k-1}>\star<n_{k-1}|L|n_{k-2}>
\star\cdots\star
<n_1|L|0>,
\label{800}
\end{eqnarray}
where $<m|L|n>$ are the matrix elements of the operator $L$ (\ref{160}).

It is not difficult to show that for a differential operator $L$ of second order, the
matrix elements $<m|L|n>$ do not vanish only for $n\le m+2$. Therefore,
the sum (\ref{800}) always contains only a {\it finite} 
number of terms, i.e., the summation
over $n_i$ is limited from above
\be
n_1\ge 0, \qquad n_i\le n_{i+1}+2, \qquad (i=1, \dots, k-1; \ n_k\equiv n).
\label{900}
\ee

\subsection{Matrix elements $<m|L|n>$}

Thus we reduced the problem of computation of the HMDS-coefficients $b_k$ to the
computation of the matrix elements of the operator $L$.
The matrix elements $<m|L|n>$ are symmetric tensors of the type $(m,n)$, i.e.
sections of the vector bundle $S_m^n(M)$.

The matrix elements $<n|L|m>$ of a Laplace type operator have been computed in our papers 
\cite{avr86b,avr91b}. They have the following general form
\bea
&&<m|L|m+2>=g^{-1}\vee I_{(m)}
\\[11pt]
&&<m|L|m+1>=0
\\[11pt]
&&<m|L|n>={m\choose n}I_{(n)}\vee Z_{(m-n)}
-{m\choose n-1}I_{(n-1)}\vee Y_{(m-n+1)}
\nonumber\\[11pt]
&&\qquad\qquad\qquad
+{m\choose n-2}I_{(n-2)}\vee X_{(m-n+2)},
\eea
where $g^{-1}$ is the metric on the cotangent bundle, 
$Z_{(n)}$ is a section of the vector bundle $S_n(M)\otimes \End(V)$, 
$Y_{(n)}$ is a section of the vector bundle $S^1_n(M)\otimes \End(V)$ and 
$X_{(n)}$ is a section of the vector bundle $S^2_n(M)$.
Here it is meant also that the binomial coefficient ${n\choose k}$ is equal to
zero if $k<0$ or $n<k$.

We will not present here explicit formulas for the objects $Z_{(n)}$, $Y_{(n)}$, 
and $X_{(n)}$, (they have been computed for arbitrary $n$ in the our papers \cite{avr86b,avr91b}),
but note that all these quantities are expressed
polynomially in terms of three sorts of geometric data:
\begin{itemize}
\item
symmetric tensors of type $(2,n)$, i.e.
sections of the vector bundle $S^2_n(M)$
\be
K_{(n)}=(s\nabla)^{n-2}{\rm Riem},
\ee
where ${\rm Riem}$ is the symmetrized Riemann tensor
\be
{\rm Riem}=R^{(c}{}_{(a}{}^{d)}{}_{b)}s^{ab}_{cd},
\label{1000}
\ee

\item
sections of the vector bundle $\End(V)\otimes S^1_n(M)$
\be
{\cal R}_{(n)}=(s\nabla)^{n-1}{\cal R},
\ee
where ${\cal R}$ is the curvature
of the connection on the vector bundle $V$ in the form
\be
{\cal R}={\cal R}^a{}_b s^a_b,
\ee

\item
$\End(V)$-valued symmetric forms, i.e.
sections of the vector bundle $\End(V)\otimes S_n(M)$, constructed from
the symmetrized covariant derivatives of the endomorphism $Q$ of the vector bundle
$V$
\be
Q_{(n)}=(s\nabla)^n Q.
\ee

\end{itemize}

From the
dimensional arguments it is obvious that the matrix elements
\hbox{$<n|L|n>$} are expressed in terms of the Riemann 
curvature tensor,
${\rm Riem}$, the bundle curvature, ${\cal R}$, and the endomorphism $Q$; 
the matrix elements $<n+1|L|n>$ --- 
in terms of the quantities $\nabla {\rm Riem}$, $\nabla{\cal R}$ and $\nabla Q$; 
the elements $<n+2|L|n>$ --- 
in terms of the quantities of the form 
$\nabla\nabla {\rm Riem}$, ${\rm Riem}\cdot {\rm Riem}$, 
etc.

\subsection{Diagramatic technique}

In the computation of the HMDS-coefficients by means of the matrix algorithm
a ``diagrammatic'' technique, i.e., a graphic method for
enumerating the different terms of the sum (\ref{800}), turns out to be very
convenient and pictorial \cite{avr86b}.

The matrix elements $<m|L|n>$ are presented by some blocks with $m$
lines coming in from the left and $n$ lines going out to the right (Fig. 1),

\begin{center}
\unitlength1mm
\begin{picture}(41,10)
\put(-2.0,-1.0){$m\,\Biggl\{$}
\put(5.0,4.0){\line(1,0){12}}
\put(7.0,0.0){$\vdots$}
\put(5.0,-1.5){\line(1,0){10}}
\put(5.0,-4.0){\line(1,0){12}}
\put(20.0,0){\circle{10}}
\put(23.0,4.0){\line(1,0){12}}
\put(32.0,0.0){$\vdots$}
\put(25.0,-1.5){\line(1,0){10}}
\put(23.0,-4.0){\line(1,0){12}}
\put(36.0,-1.0){$\Biggl\}\,n$}
\end{picture}
\end{center}
\vglue4mm
\begin{center}
Fig. 1
\end{center}
and the product of the matrix elements $<m|L|k>\star<k|L|n>$ --- by two blocks
connected by $k$ intermediate lines (Fig. 2),

\begin{center}
\unitlength1mm
\begin{picture}(61,10)
\put(-2.0,-1.0){$m\,\Biggl\{$}
\put(5.0,4.0){\line(1,0){12}}
\put(7.0,0.0){$\vdots$}
\put(5.0,-1.5){\line(1,0){10}}
\put(5.0,-4.0){\line(1,0){12}}
\put(20.0,0){\circle{10}}
\put(23.0,4.0){\line(1,0){14}}
\put(25.0,-1.5){\line(1,0){10}}
\put(23.0,-4.0){\line(1,0){14}}
\put(26.0,-1.0){$k\,\Biggl\{$}
\put(32.0,0.0){$\vdots$}
\put(40.0,0){\circle{10}}
\put(43.0,4.0){\line(1,0){12}}
\put(52.0,0.0){$\vdots$}
\put(45.0,-1.5){\line(1,0){10}}
\put(43.0,-4.0){\line(1,0){12}}
\put(56.0,-1.0){$\Biggl\}\,n$}
\end{picture}
\end{center}
\vglue4mm
\begin{center}
Fig. 2
\end{center}
that represents the contractions of the corresponding tensor indices (the inner product).

To obtain the coefficient $<n|b_k>$ one should draw, first, all possible
diagrams which have $n$ lines incoming from the left and which are
constructed from $k$ blocks connected in all possible ways by any number of
intermediate lines. 
When doing this, one should keep in mind that the number of
the lines, going out of any block, cannot be greater than the number of the lines,
coming in, by more than two and by exactly one. 
Then one should
sum up all diagrams with the weight determined for each diagram by the number
of intermediate lines from the analytical formula (\ref{800}). 
Drawing of such diagrams is of no difficulties. 
This helps to keep under control the whole variety of different terms.
Therefore, the main problem is reduced to the computation of some standard blocks,
which can be computed once and for all. 

For example, the diagrams for the diagonal values of the HMDS-coefficients
\hbox{$[b_k]=<0|b_k>$} have the form,

\unitlength1mm
\newsavebox{\bone}
\savebox{\bone}(22,1.5)[l]
{\multiput(4,2)(7,0){1}{\circle{4}}
\put(19.35,3.5){}}
\be
[b_1]=\usebox{\bone}
\ee

\unitlength1mm
\newsavebox{\btwoa}
\savebox{\btwoa}(22,1.5)[l]
{\multiput(4,2)(7,0){2}{\circle{4}}
\put(19.35,3.5){}}
\newsavebox{\btwob}
\savebox{\btwob}(22,1.5)[l]
{\multiput(4,2)(7,0){2}{\circle{4}}
\put(5.3,3.5){\line(1,0){4.37}}
\put(5.3,0.5){\line(1,0){4.37}}}
\be
[b_2]=\usebox{\btwoa}
\!\!\!\!\!\!\!\!\!\!
+{1\over 3}\usebox{\btwob}
\ee

%
%

\unitlength1mm
\newsavebox{\fblocka}
\savebox{\fblocka}(22,1.5)[l]
{\multiput(4,2)(7,0){3}{\circle{4}}
\put(19.35,3.5){}}
\newsavebox{\fblockb}
\savebox{\fblockb}(22,1.5)[l]
{\multiput(4,2)(7,0){3}{\circle{4}}
\put(12.4,3.5){\line(1,0){4.37}}
\put(12.4,0.5){\line(1,0){4.37}}}
\newsavebox{\fblockc}
\savebox{\fblockc}(22,1.5)[l]
{\multiput(4,2)(7,0){3}{\circle{4}}
\put(5.3,3.5){\line(1,0){4.37}}
\put(5.3,0.5){\line(1,0){4.37}}}
\newsavebox{\fblockd}
\savebox{\fblockd}(22,1.5)[l]
{\multiput(4,2)(7,0){3}{\circle{4}}
\put(5.3,3.5){\line(1,0){4.37}}
\put(5.3,0.5){\line(1,0){4.37}}
\put(12.9,2){\line(1,0){3.1}}}
\newsavebox{\fblocke}
\savebox{\fblocke}(22,1.5)[l]
{\multiput(4,2)(7,0){3}{\circle{4}}
\put(5.3,3.5){\line(1,0){4.37}}
\put(5.3,0.5){\line(1,0){4.37}}
\put(12.4,3.5){\line(1,0){4.37}}
\put(12.4,0.5){\line(1,0){4.37}}}
\newsavebox{\fblockf}
\savebox{\fblockf}(22,1.5)[l]
{\multiput(4,2)(7,0){3}{\circle{4}}
\put(5.3,3.5){\line(1,0){4.37}}
\put(5.3,0.5){\line(1,0){4.37}}
\put(12.35,3.5){\line(1,0){4.37}}
\put(12.35,0.5){\line(1,0){4.37}}
\put(12.8,2.5){\line(1,0){3.2}}
\put(12.8,1.5){\line(1,0){3.2}}}
\begin{eqnarray}
[b_3]&=&\usebox{\fblocka}
+{1\over 3}\usebox{\fblockb}
+{2\over 4}\usebox{\fblockc}
\nonumber\\[10pt]
& &
+{2\over 4}\cdot{1\over 2}\usebox{\fblockd}
+{2\over 4}\cdot{1\over 3}\usebox{\fblocke}
+{2\over 4}\cdot{1\over 5}\usebox{\fblockf},
\nonumber\\[10pt]
& &
\end{eqnarray}

\subsection{Remarks}

Let us make some remarks about the elaborated technique.
\begin{itemize}
\item
this technique is applicable for a {\it generic} Riemannian manifold $M$
and for a {\it generic} vector bundle $V$ of {\it arbitrary} dimensions,

\item
this technique is manifestly {\it covariant}, which is an inestimable advantage in quantum
field theory, especially in quantum gravity and gauge theories,

\item
since it is purely local, it is also valid for manifolds with boundary and noncompact
manifolds, provided one considers the local HMDS-coefficients $b_k(x,x')$ 
in a small neighbourhood
$B$ of the diagonal of $M\times M$ that does not intersect with the boundary,
$B\cap \partial M=\emptyset$,

\item
moreover, this technique works also in the case of pseudo-Riemannian manifolds
and {\it hyperbolic} differential operators,

\item
this method is {\it direct}---it does not need to use any sophisticated functorial properties
of the integrated coefficients $B_k$,

\item
it gives not only the diagonal values of the HMDS-coefficients $[b_k]$ but also the diagonal
values of all their derivatives; thus it gives immediately the asymptotics of the trace of
derivatives of the heat kernel
\be
\Tr_{L^2}P\exp(-tF),
\ee
where $P$ is a differential operator,

\item
due to use of {\it symmetric} forms and {\it symmetric} covariant derivatives the 
famous `{\it combinatorial explosion}' in the complexity of the HMDS-coefficients 
is avoided,

\item
the developed technique is very algorithmic and well suited to {\it 
automated computation} ---
there are a number of usual algebraic operations on symmetric tensors that seems to be
easily programmed, the needed input, i.e. the matrix elements $<n|L|m>$, is computed 
in advance analytically and is already known,

\item
the developed method is {\it very powerful}; it enabled us to compute for the first time
the diagonal value of the fourth HMDS-coefficient $[b_4]$
\cite{avr86b,avr91b}.
(The third coefficient $[b_3]$ was computed first by {\sc Gilkey} 
\cite{gilkey75b}.)

\item
last, this technique enables one not only to carry out explicit computations, but
also to analyse the general structure of the HMDS-coefficients $b_k$ for 
all orders $k$.

\end{itemize}

\section{Covariant approximation schemes for the heat kernel}

\subsection{General structure of HMDS-coefficients}

Now we are going to investigate the general structure of the HMDS-coefficients.
We will follow mainly our papers \cite{avr86b,avr91b,avr93b,avr95b,avr94a,avr96a} (see also our review papers 
\cite{avr95c,avr94b,avr95d}).
Our analysis will be again {\it purely local}. 
Since locally one can always expand the metric,
the connection and the endomorphism $Q$ in the covariant Taylor series, 
they are completely characterized by their Taylor coefficients, i.e.
the covariant derivatives of the curvatures, more precisely by the objects
$K_{(n)}$, ${\cal R}_{(n)}$ and $Q_{(n)}$ introduced above.
We introduce the following notation for all of them
\be
\Re_{(n)}=\{K_{(n+2)}, {\cal R}_{(n+1)}, Q_{(n)}\}, 
\qquad (n=0,1,2,\dots),
\ee
and call these objects {\it covariant jets}. $n$ will be called the {\it order}
of a jet $\Re_{(n)}$. 
Further we introduce an infinite set of covariant jets of all orders
\be
{\cal J}=\{\Re_{(n)};\ (n=0,1,2,\dots)\}.
\ee

The first two HMDS-coefficients have a very well known form \cite{gilkey75b,avr91b}
\be
B_{0}=\int\limits_M d\vol(x)\tr_V I.
\ee
\be
B_1=\int\limits_M d\vol(x)\tr_V \left(Q-{1\over 6}R\right),
\ee 
where $R$ is the scalar curvature. 

As far as the higher order coefficients $B_k$, $(k\ge 2)$, are concerned 
they are integrals of local invariants which are polynomial in the jets \cite{gilkey95}.
One can classify all the terms in them according to the number of
the jets and their order.
The terms linear in the jets in higher order coefficients
$B_k$, ($k\ge 2$), are given by integrals of total derivatives,
symbolically $\int_Md\vol(x)\,\tr_V\,\sq^{k-1}\Re$. They are calculated explicitly 
in \cite{avr86b}.
Since the total derivative do not contribute to an integral 
over a complete compact manifold,
it is clear that the linear terms vanish. 
Thus $B_k$, $(k=2,3,\dots)$, begin with the terms quadratic in the jets.
These terms contain the jets of highest order (or the {\it leading derivatives} of the curvatures)
and can be shown to be of the form $\int_Md\vol(x) \tr_V\,\Re \sq^{k-2} \Re$. 
Then it follows a class of terms cubic in the jets etc.. 
The last class of terms does not contain any 
covariant derivatives at all but only the powers of the curvatures. 
In other words, the higher order HMDS-coefficients have a general structure,
which can be presented symbolically in the form
\bea
B_k &=&\int\limits_M d\vol(x)\,{\rm tr}_V\,\Biggl\{\Re \sq^{k-2} \Re
+\sum_{0\le i\le 2k-6}\Re\ \na^i\Re\ \na^{2k-6-i}\Re
\nonumber\\
&&+\cdots+\sum_{0\le i\le k-3}\Re^i(\na \Re)\Re^{k-i-3}(\na\Re)
+\Re^k\Biggr\}.
\eea

\subsection{Leading derivatives in heat kernel asymptotics}

More precisely,
all quadratic terms can be reduced to a finite number
of invariant structures, viz. \cite{avr86b,avr91b} 
\begin{eqnarray}
B_{k,2}&=&
{k!(k-2)!\over 2(2k-3)!}
\int\limits_M d\vol(x)\,{\rm tr}_V\,
\Biggl\{f^{(1)}_kQ\Square^{k-2}Q
+2f^{(2)}_k{\cal R}^{\beta\mu}\nabla_\beta
\Square^{k-3}\nabla_\alpha{\cal R}^\alpha{}_\mu
\nonumber\\[10pt]
&&
+f^{(3)}_k Q\Square^{k-2}R
+f^{(4)}_kR_{\mu\nu}\Square^{k-2}R^{\mu\nu}
+f^{(5)}_kR\Square^{k-2}R
\Biggr\},
\label{4-5-3.53}
\end{eqnarray}
where $R_{\mu\nu}$ is the Ricci tensor and $f^{(i)}_k$ are some numerical coefficients.
These numerical coefficients can be computed by the technique developed in the 
previous section.
From the formula
(\ref{800}) we have for the diagonal coefficients $[b_k]$ up to cubic terms in the jets 
\begin{eqnarray}
[b_k]&=&<0|b_k>={(-1)^{k-1}\over {2k-1\choose k}}<0;k-1|L|0>
\nonumber\\[10pt]
&&
+(-1)^k\sum\limits_{i=1}^{k-1}\,\sum\limits_{n_i=0}^{2(k-i-1)}
{{2k-1\choose i}\over {2k-1\choose k}{2i+n_i-1\choose i}}
<0;k-i-1|L|n_i><n_i;i-1|L|0>
\nonumber\\[11pt]
&&
+O(\Re^3),
\label{4-5-3.52}
\end{eqnarray}
where 
\be
<n;k|L|m>=(\vee^k g^{-1})\star<n|L|m>
\ee
and $O(\Re^3)$ denote terms of third order in the jets.

By computing the matrix elements in the second order in the jets and
integrating over $M$ one obtains \cite{avr86b,avr91b}
\bea
f^{(1)}_k&=&1\\
f^{(2)}_k&=&{{1}\over {2(2k-1)}}\\
f^{(3)}_k&=&{{k-1}\over {2(2k-1)}}\\
f^{(4)}_k&=&{{1}\over {2(4k^2-1)}}\\
f^{(5)}_k&=&{{k^2-k-1}\over {4(4k^2-1)}}.
\label{4-5-4.7}
\eea

One should note that the same results were obtained by a completely different
method by {\sc Branson, Gilkey and \O rsted} \cite{branson90b}. 

\subsection{`Summation' of asymptotic expansion}

Let us consider the situation when the curvatures are small but rapidly varying, i.e.
the derivatives of the curvatures are more important than the powers of them.
Then the leading derivative terms in the heat kernel are the largest ones.
Thus the trace of the heat kernel has the form
\be
\Tr_{L^2}\exp(-tF)\sim 
(4\pi t)^{-d/2}\left\{B_0-tB_1+{t^2\over 2}H_2(t)\right\}+O(\Re^3),
\ee
where $H_2(t)$ is some complicated {\it nonlocal} functional
that has the following asymptotic expansion as $t\to 0$
\be
H_2(t)\sim2\sum_{k\ge 2}{(-t)^{k-2}\over k!}B_{k,2}+O(\Re^3).
\ee
Using the results for $B_{k,2}$ one can easily construct such a functional
$H_2$ just by a formal summing the leading derivatives 
\bea
H_2&=&\int\limits_M d\vol(x)\,\tr_V\,\Biggl\{
Q\gamma^{(1)}(-t\Square)Q
+2{\cal R}_{\alpha\mu}\nabla^\alpha{{1}\over {\Square}}
\gamma^{(2)}(-t\Square)\nabla_\nu{\cal R}^{\nu\mu}
\nonumber\\
&&-2Q\gamma^{(3)}(-t\Square)R 
+R_{\mu\nu}\gamma^{(4)}(-t\Square)R^{\mu\nu}
+R\gamma^{(5)}(-t\Square)R
\Biggr\},
\label{4-5-4.13}
\eea
where $\gamma^{(i)}(z)$ are entire functions defined by 
\cite{avr86b,avr91b}
\be
\gamma^{(i)}(z)=\sum\limits_{k\ge 0}{k!\over (2k+1)!}f^{(i)}_kz^k
=\int\limits_0^1 d\xi\, f^{(i)}(\xi)
\exp\left(-{{1-\xi^2}\over {4}}z\right)
\label{4-5-4.12}
\ee
where 
\bea
f^{(1)}(\xi)&=&1\\
f^{(2)}(\xi)&=&{{1}\over {2}}\xi^2\\
f^{(3)}(\xi)&=&{{1}\over {4}}(1-\xi^2)\\
f^{(4)}(\xi)&=&{{1}\over {6}}\xi^4\\
f^{(5)}(\xi)&=&{{1}\over {48}}(3-6\xi^2-\xi^4).
\label{4-5-4.9}
\eea
Therefore, $H_2(t)$ can be regarded as {\it generating functional} for quadratic terms
$B_{k,2}$ (leading derivative terms) in all HMDS-coefficients $B_k$.
It plays also a very important role in investigating the nonlocal structure of the effective
action in quantum field theory in high-energy approximation \cite{avr91b}.

Let us note also that the function $B_q$ introduced in the Sect. 1 can be obtained 
just by analytical continuation of the formula for $B_k$ from integer points $k$ to a
complex plane $q$.

\subsection{Covariantly constant background}

Let us consider now the opposite case, when the curvatures are strong but slowly
varying, i.e. the powers of the curvatures are more important than the derivatives of them.
The main terms in this approximation are the terms without any covariant derivatives
of the curvatures, i.e. the lowest order jets.
We will consider mostly the zeroth order of this approximation
which corresponds simply to covariantly constant background curvatures
\be
\na {\rm Riem} = 0,\qquad \na {\cal R}=0,
\qquad \na Q = 0.
\ee

The asymptotic expansion of the trace of the heat kernel
\be
\Tr_{L^2}\exp(-tF)\sim (4\pi t)^{-d/2}\sum_{k\ge 0}{(-t)^k\over k!}B_{k}.
\ee
determines then {\it all} the terms without covariant derivatives 
(highest order terms in the jets), $B_{k,k}$, in all HMDS-coefficients $B_k$.
These terms do not contain any covariant derivatives and are just
polynomials in the curvatures and the endomorphism $Q$. 
Thus the trace of the heat kernel is a generating
functional for all HMDS-coefficients for a covariantly constant background, in
particular, for all symmetric spaces.
Thus the problem is to calculate  the trace of the heat kernel
for covariantly constant background.

\subsubsection{Algebraic approach}

There exist a very elegant indirect way to construct the heat 
kernel without solving the heat equation but using only the 
commutation relations of some covariant first order differential 
operators \cite{avr93b,avr95b,avr94a,avr96a}.
The main idea is in a generalization of the usual Fourier 
transform to the case of operators and consists in the following.

Let us consider for a moment a trivial case, where the 
curvatures vanish but not the potential term:
\be
{\rm Riem}= 0,\qquad {\cal R}=0,\qquad \nabla Q=0.
\ee
In this case the operators of covariant derivatives obviously 
commute and form together with the potential term an Abelian 
Lie algebra
\be
[\na_\mu,\na_\nu]=0,\qquad [\na_\mu, Q]=0. 
\ee
It is easy to show that the {\it heat semigroup operator} can 
be presented in the form
\be
\exp(-t F)=(4\pi t)^{-d/2}\exp(-tQ)\int\limits_{\RR^d} d\vol(k)
\exp\left(-{1\over 4t}<k,g k>
+k\cdot\na\right),
\label{1300}
\ee
where $<k,g k>=k^\mu g_{\mu\nu}k^\nu$, $k\cdot\nabla=k^\mu \nabla_\mu$.
Here, of course, it is assumed that the covariant derivatives 
commute also with the metric
\be
[\nabla, g]=0. 
\ee
Acting with this operator on the Dirac distribution and using the 
obvious relation
\be
\exp(k\cdot\na)\delta(x,x')\big\vert_{x=x'}=\delta(k),
\ee
one integrates easily over $k$ and obtains the trace of the heat kernel 
\be
\Tr_{L^2}\exp(-tF)=(4\pi t)^{-d/2}\int\limits_M d\vol(x)\tr_V\exp(-tQ).
\ee

In fact, the covariant differential operators 
$\nabla$ do not commute, their commutators being proportional to the 
curvatures $\Re$. 
The commutators of covariant derivatives $\na$ with the curvatures 
$\Re$ give the first 
derivatives of the curvatures, i.e. the jets $\Re_{(1)}$, the 
commutators of 
 covariant derivatives with $\Re_{(1)}$ give the second jets 
 $\Re_{(2)}$, etc.
Thus the operators $\nabla$ together with the 
whole set of the jets ${\cal J}$ form an {\it infinite} dimensional 
Lie algebra 
${\cal G}=\{\nabla, \Re_{(i)}; (i=1,2,\dots)\}$.

To evaluate the heat kernel in the considered (low-energy) approximation one 
can take into account a {\it finite} number of low-order jets, 
i.e. the 
low-order covariant derivatives of the background fields,
$\{\Re_{(i)}; (i\le N)\}$,  and neglect all 
the higher order jets, i.e. the covariant derivatives of 
higher orders,
i.e. put $\Re_{(i)}=0$\ for $i> N$.
Then one can show that there exist a set of covariant differential 
operators 
that together with the background fields and their low-order 
derivatives 
generate a {\it finite} dimensional Lie algebra 
${\cal G}_N=\{\nabla, \Re_{(i)};
(i=1,2,\dots,N)\}$ \cite{avr95c,avr94b,avr95d}.

Thus one can try to generalize 
the above idea in such a way that (\ref{1300}) would be the zeroth 
approximation 
in the commutators of the covariant derivatives, i.e. in the 
curvatures. 
Roughly speaking, we are going to find a representation of the 
heat semigroup 
{\it operator} 
in the form
\be
\exp(-t F)=
\int\limits_{\RR^D} dk\,
\Phi(t,k)\exp\left(-{1\over 4t}<k,\Psi(t) k>
+k\cdot T\right)
\ee
where $<k,\Psi(t) k>=k^A\Psi_{AB}(t)k^B$, 
$k\cdot T=k^A T_A$, ($A=1,2,\dots,D$), 
 $T_A=X^\mu_A\na_\mu+Y_A$ are some first order differential operators 
 and the functions $\Psi(t)$ and $\Phi(t,k)$ are expressed in 
 terms  of commutators of these 
operators--- i.e., in terms of the curvatures.

In general, the operators $T_A$ do not form a closed finite 
dimensional algebra because at each step taking more 
commutators there appear more and more derivatives of the 
 curvatures. It is the {\it low-energy reduction} 
${\cal G}\to {\cal G}_N$, i.e. the restriction to the 
low-order jets, that actually closes the algebra ${\cal G}$ 
of the operators $T_A$ and the background jets, i.e. makes 
it finite dimensional.

Using this representation one can, as above, act with 
$\exp(k\cdot T)$ on the Dirac distribution to get the 
heat kernel. The main point of this idea is that it is 
much easier to calculate the action of the exponential 
of the {\it first} order operator $k\cdot T$ on the Dirac
distribution
than that of the exponential of the second 
order operator $\sq$.

\subsubsection{Covariantly constant bundle curvature and
covariantly constant
endomorphsim $Q$ in flat space}

Let us consider now the more complicated case of nontrivial 
covariantly constant curvature of the connection on the vector bundle $V$
in flat space:
\be
{\rm Riem}= 0,\qquad \nabla{\cal R}=0,
\qquad \nabla Q=0.
\ee

Using the condition of covariant constancy of the curvatures
one can show that in this case the covariant derivatives form 
a {\it nilpotent} Lie algebra \cite{avr93b}
\bea
&&[\nabla_\mu,\nabla_\nu]={\cal R}_{\mu\nu}, \\
&&[\nabla_\mu,{\cal R}_{\alpha\beta}]=[\nabla_\mu,Q]=0,\\
&&[\h R_{\mu\nu},\h R_{\a\b}]=[\h R_{\mu\n},Q]=0.
\eea

For this algebra one can prove a theorem expressing the heat 
semigroup operator in terms of an average over the 
corresponding Lie group
\cite{avr93b}
\bea
\exp(-t F)&=&(4\pi t)^{-d/2}\exp(-tQ)
\det^{1/2}_{\End(TM)}\left({t\h R\over \sinh(t\h R)}\right)\\
&&\times\int\limits_{\RR^d}d\vol(k)
\exp\left(-{1\over 4t}<k, g t\h R \coth(t\h R) k>
+k\cdot\nabla\right),
\eea
where $k\cdot\nabla=k^\mu\nabla_\mu$. 
Here functions of the curvatures $\h R$ are understood as 
functions of sections of the bundle $\End(TM)\otimes\End(V)$,
and the determinant 
$\det_{\End(TM)}$ is taken with respect to $\End(TM)$ indices, 
$\End(V)$ indices being intact.

It is not difficult to show that also in this case we have
\be
\exp(k\cdot\nabla)\d(x,x')\big\vert_{x=x'}
=\d(k).
\ee
Subsequently, the integral over $k^\mu$ becomes trivial and 
we obtain immediately the trace of the heat kernel \cite{avr93b}
\be
\Tr_{L^2}\exp(-tF)=(4\pi t)^{-d/2}\int\limits_M d\vol(x)\tr_V\exp(-tQ)
\det^{1/2}_{\End(TM)}\left({t\h R \over \sinh(t\h R)}\right). 
\ee
Expanding it in a power series in $t$ one can find {\it all} 
covariantly constant terms in {\it all} HMDS-coefficients $B_k$.

As we have seen the contribution of the bundle curvature 
$\h R_{\mu\n}$ is not as trivial as that of the potential term. 
However, the algebraic approach does work in this case too. 
It is a good example how one can get the heat kernel without 
solving any differential equations but using only the algebraic 
properties of the covariant derivatives.

\subsubsection{Contribution of two first derivatives of 
the endomorphism $Q$}

In fact, in flat space it is possible to do a bit more, i.e. 
to calculate the contribution of the first and the second 
derivatives of the potential term $Q$
\cite{avr95b}.
That is we consider the case when the derivatives of the 
endomorphism $Q$ vanish only starting from the {\it third} 
order, i.e.
\be
{\rm Riem}= 0,\qquad \nabla{\cal R}=0,\qquad 
\nabla\nabla\nabla Q=0.
\ee
Besides we assume the background to be {\it Abelian}, 
i.e. all the nonvanishing background quantities, 
$\h R_{\a\b}$, $Q$, $Q_{;\mu}\equiv\nabla_\mu Q$ and 
$Q_{;\n\mu}\equiv \nabla_\n\nabla_\mu Q$, commute with each other. 
Thus we have again a nilpotent Lie algebra
\bea
&&[\nabla_\mu, \nabla_\nu]={\cal R}_{\mu\nu}\\
&&[\nabla_\mu,Q]=Q_{;\m}\\
&&[\nabla_\m,Q_{;\n}]=Q_{;\n\m}
\label{1400}
\eea
all other commutators being zero.

Now, let us represent the endomorphism $Q$ in the form
\be
Q=\Omega-\a^{ik}N_iN_k,
\ee
where $(i=1,\dots,q; q\le d),$ $\a^{ik}$ is some constant 
symmetric nondegenerate $q\times q$ matrix, $\Omega$ is a 
covariantly constant endomorphism and $N_i$ are some endomorphisms with 
vanishing second covariant derivative:
\be
\nabla \Omega=0, \qquad \nabla\nabla N_i=0.
\ee
Next, let us introduce the operators $X_A=(\nabla_\m, N_i)$, 
$(A=1,\dots, d+q)$ and the matrix
\be
(\h F_{AB})=\left(\matrix{
\R_{\m\n} & N_{i;\m}\cr
-N_{k;\n} & 0       \cr}\right),
\ee
with $N_{i;\m}\equiv \nabla_\m N_i$.

The operator $F$ can now be written in the form
\be
F=-G^{AB}X_A X_B+\Omega,
\ee
where
\be
(G^{AB})=\left(\matrix{g^{\m\n} & 0 \cr
			0 & \a^{ik} \cr}\right).
\ee
and the commutation relations (\ref{1400}) take a more compact form
\be
[X_A, X_B]=\h F_{AB}
\ee
all other commutators being zero.

This algebra is again a nilpotent Lie algebra.
Thus one can apply the previous theorem  in this case too to get
\cite{avr95b}
\bea
\exp(-t F)&=&(4\pi t)^{-{(d+q)}/2}\exp(-t\Omega)
\det^{1/2}\left({t\h F\over \sinh(t\h F)}\right)
\nonumber\\
&&
\times\int\limits_{\RR^{d+q}} d k G^{1/2}
\exp\left(-{1\over 4t}<k, G t\h F \coth(t\h F)k>
+k\cdot X\right),
\eea
where $G=\det G_{AB}$ and $k\cdot X=k^A X_A$.

Thus we have expressed the heat semigroup operator in terms 
of the operator 
\hbox{$\exp(k\cdot X)$}. The integration over $k$ is 
Gaussian except for the noncommutative part.
Splitting the integration variables $(k^A)=(q^\m, \om^i)$ and 
using the 
Campbell-Hausdorf formula we obtain \cite{avr95b}
\be
\exp(k\cdot X)\d(x,x')\Big\vert_{x=x'}
=\exp(\om\cdot N)\d(q),
\ee
where $\om\cdot N=\om^i N_i$. 

Further, after taking off the trivial 
integration over $q$
and a Gaussian integral over $\om$,  
we obtain the trace of the heat kernel in a very simple form
\cite{avr95b} 
\be
\Tr_{L^2}\exp(-tF)=(4\pi t)^{-d/2}\int\limits_Md\vol(x)\tr_V
\Phi(t) \exp\left[-tQ
+{1\over  4} t^3 <\nabla Q, \Psi(t)g^{-1}\nabla Q>\right],
\label{1500}
\ee
where
$
<\nabla Q, \Psi(t)g^{-1}\nabla Q>
=\nabla_\m Q\Psi^\m_{\ \n}(t)g^{\n\l}\nabla_\l Q,
$
\bea
\Phi(t)&=&\det_{\End(TM)}^{-1/2}K(t)\,
\det^{-1/2}_{\End(TM)}\left\{1+t^2[E(t)-S(t)K^{-1}(t)S(t)]P\right\}\,
\nonumber\\[11pt]
&&\times
\det^{-1/2}_{\End(TM)}\left[1+t^2C(t)P\right],
\eea
\be
\Psi(t)=\{\Psi^\m_{\ \n}(t)\}=\left[1+t^2C(t)P\right]^{-1}C(t),
\ee
$P$ is the matrix determined by second derivatives of the 
potential term,
\be
P=\left\{P^\m_{\ \n}\right\}, \qquad 
P^\m_{\ \n}={1\over 2} g^{\m\l}\nabla_\n \nabla_\l Q,
\ee
 and
the matrices $C(t)=\{C^\m_{\ \n}(t)\}$, $K(t)=\{K^\m_{\ \n}(t)\}$ 
$S(t)=\{S^\m_{\ \n}(t)\}$ and $E(t)=\{E^\m_{\ \n}(t)\}$ are defined by 
\be
C(t)=\oint\limits_C{dz\over 2\pi i}\,t\,
\coth({tz^{-1}})(1-z\h R-z^2 P)^{-1},
\ee
\be
K(t)=\oint\limits_C{dz\over 2\pi i}{t\over z^{2}}\sinh(tz^{-1})(1-z\h R-z^2 P)^{-1},
\ee
\be
S(t)=\oint\limits_C{dz\over 2\pi i}\,{t\over z}\,\sinh(tz^{-1})(1-z\h R-z^2 P)^{-1},
\ee
\be
E(t)=\oint\limits_C{dz\over 2\pi i}\,t\,\sinh(tz^{-1})(1-z\h R-z^2 P)^{-1},
\ee
where the integral is taken along a sufficiently small closed contour $C$ 
that encircles the origin counter-clockwise, so that $F(z)=(1-z\h R-z^2 P)^{-1}$ is analytic
 inside this contour.

The formula (\ref{1500}) exhibits the general structure of the trace of the heat 
kernel.
Namely, one sees immediately how the endomorphism $Q$ and its 
first derivatives $\nabla Q$ enter the result. The nontrivial information is 
contained only in a 
scalar, $\Phi(t)$, and a tensor, $\Psi_{\m\n}(t)$, functions 
which are 
constructed purely from the curvature $\R_{\m\n}$ and the 
{\it second} derivatives of the endomorphism $Q$, $\nabla\nabla Q$. 

So, we 
conclude that the HMDS-coefficients $B_k$ are 
constructed from three different types of scalar (connected) blocks, 
$Q$, $\Phi_{(n)}(\R, \nabla\nabla Q)$ and 
$\nabla_\m Q\Psi^{\m\n}_{(n)}(\R, \nabla\nabla Q)\nabla_\n Q$. 
They are listed explicitly up to $B_8$ in \cite{avr95c}.  

\subsubsection{Symmetric spaces}

Let us now generalize the algebraic approach
to the case of the {\it curved} manifolds with covariantly constant 
Riemann curvature and the trivial bundle connection
\cite{avr94a,avr96a}
\be
\nabla {\rm Riem}=0, \qquad \R=0, \qquad \nabla Q=0.
\label{1600}
\ee

First of all, we give some definitions.
The condition (\ref{1600}) defines, as we already said above, 
the geometry of {\it locally symmetric spaces}. 
A Riemannian locally symmetric space which is simply connected and 
complete is
{\it globally symmetric space} (or, simply,  symmetric space).
A symmetric space is said to be of {\it compact, noncompact 
{\rm or} 
Euclidean type} if {\it all} sectional curvatures 
$K(u,v)=R_{abcd}u^av^bu^cv^d$ 
are positive, negative or zero. 
A direct product of symmetric 
spaces of 
compact and noncompact types is called {\it semisimple} 
symmetric space. 
A generic complete simply connected 
Riemannian symmetric space is a direct product of a flat space and a 
semisimple symmetric space. 

It should be noted that our analysis in this paper is purely 
{\it local}.  We are looking for a 
{\it universal} local function of the curvature invariants, 
that reproduces adequately  the asymptotic expansion of the 
trace of the heat kernel.
This function should give {\it all} the terms without covariant 
derivatives of the curvature in the
asymptotic expansion of the heat kernel, i.e. 
in other words {\it all} HMDS-coefficients $B_k$  for {\it any}
locally symmetric space.

It is well known that the HMDS-coefficients have a {\it universal} 
structure, i.e.
they are polynomials in the background jets (just in curvatures 
in case of symmetric spaces) 
with the numerical coefficients that do not depend
on the global properties of the manifold, on the dimension, on 
the signature of the metric etc. It is this universal structure we 
are going to study.

It is obvious that any flat subspaces do not contribute to 
the HMDS-coefficients $B_k$. 
Therefore, to find this universal structure it is sufficient 
to consider only semisimple 
symmetric spaces. 
Moreover, since HMDS-coefficients are 
analytic in the curvatures, one can restrict oneself only 
to symmetric spaces of {\it compact} type.
Using the factorization property of the heat kernel and 
the duality 
between compact and noncompact symmetric spaces one can 
obtain then the 
 results for the general case by analytical continuation.
That is why we consider only the case of {\it compact} symmetric 
spaces  when the sectional curvatures and the metric
are {\it positive} definite. 

Let $e_a$ be a basis in the tangent bundle 
which is {\it covariantly constant (parallel)} 
along the geodesic.
The frame components of the curvature tensor of a symmetric 
space are, obviously, constant and can be presented in the form 
\be
R_{abcd} = \b_{ik}E^i_{\ ab}E^k_{\ cd}, 
\ee
where $E^i_{ab}$, $(i=1,\dots, p; p \le d(d-1)/2)$, is some 
set of antisymmetric matrices and $\b_{ik}$ is some symmetric 
nondegenerate $p\times p$ matrix.
The traceless matrices $D_i=\{D^a_{\ ib}\}$ defined by
\be
D^a_{\ ib}=-\b_{ik}E^k_{\ cb}g^{ca}= - D^a_{\ bi} 
\ee
are known to be the generators of the {\it holonomy algebra} 
${\cal H}$
\be
[D_i, D_k] = F^j_{\ ik} D_j, 
\ee
where $F^j_{\ ik}$ are the structure constants. 

In symmetric spaces a much richer algebraic structure exists.
Indeed, 
let us define the quantities $C^A_{\ BC}=-C^A_{\ CB}$, 
$(A=1,\dots, D; D=d+p)$:
\be
C^i_{\ ab}=E^i_{\ ab}, \quad C^a_{\ ib}
=D^a_{\ ib}, \quad C^i_{\ kl}=F^i_{\ kl}, 
\ee
\be
C^a_{\ bc}=C^i_{\ ka}=C^a_{\ ik}=0,
\ee
and the matrices $C_A=\{C^B_{\ AC}\}=(C_a,C_i)$:
\be
C_a = \left( \matrix{ 0          & D^b_{\ ai}   \cr
		      E^j_{\ ac} & 0            \cr}\right), 
\qquad
C_i = \left( \matrix{ D^b_{\ ia} & 0            \cr
		      0          & F^j_{\ ik}   \cr}\right).
\ee
One can show that they satisfy the Jacobi identities \cite{avr94a,avr96a}
\be
[C_A, C_B]=C^C_{\ AB}C_C
\ee
and, hence, define a Lie algebra ${\cal G}$ of dimension $D$ with 
the structure constants $C^A_{\ BC}$, the matrices
$C_A$ being the generators of adjoint representation.

In symmetric spaces one can find explicitly the generators 
of the infinitesimal isometries, i.e. the Killing
vector fields $\xi_A$,
and show that they form a Lie algebra of isometries that is 
(in case of semisimple symmetric space) isomorphic to the
Lie algebra ${\cal G}$,
viz.
\be
[\xi_A,\xi_B]=C^C_{\ AB}\xi_C.    
\ee
Moreover, introducing a symmetric nondegenerate $D\times D$ 
matrix
\be
\g_{AB} = \left(\matrix{ g_{ab} & 0             \cr
	       0                & \b_{ik}        \cr}\right),
\ee
that plays the role of the metric on the algebra ${\cal G}$, 
one can
express the operator $F$ in semisimple symmetric spaces in
terms of the generators of isometries
\be
F=-\g^{AB}\xi_A\xi_B+Q,
\ee
where $\g^{AB}=(\g_{AB})^{-1}$.

Using this representation one can prove a theorem that 
presents the heat 
semigroup operator in terms of some average over the 
group of isometries $G$ \cite{avr94a,avr96a}
\bea
\exp(-t F)&=&
(4\pi t)^{-D/2}\exp\left[-t\left(Q-{1\over 6} R_G\right)\right]
\label{1700}\\[12pt]
&&
\times\int\limits_{\RR^D} d k\g^{1/2}
\det^{1/2}_{{\rm Ad}({\cal G})}\left({\sinh(k\cdot C/2)\over k\cdot C/2}\right)
\exp\left(-{1\over 4t}<k,\g k>
+k\cdot \xi\right)
\nonumber
\eea
where $\g=\det\g_{AB}$, $k\cdot C=k^A C_A$, $k\cdot \xi=k^A\xi_A$, 
and $R_G$ is the scalar curvature of the group of isometries $G$
\be
R_G= -{1\over 4}\g^{AB} C^C_{\ AD}C^D_{\ BC}. 
\ee

Acting with this operator on the Dirac distribution $\d(x,x')$ one can,
in principle, evaluate the off-diagonal heat kernel 
$\exp(-tF)\d(x,x')$, i.e. for non-coinciding points $x\ne x'$
(see \cite{avr96a}).
To calculate the trace of the heat kernel,
it is sufficient to compute only the coincidence limit $x=x'$.  
Splitting the integration variables $k^A=(q^a,\om^i)$ and 
solving the equations of characteristics one can obtain the 
action of the isometries on the Dirac distribution  \cite{avr94a,avr96a}
\be
\exp\left(k\cdot\xi\right)\d(x,x')\Big\vert_{x=x'}
=\det^{-1}_{\End(TM)}\left({\sinh(\om\cdot D/2)\over 
\om\cdot D/2}\right)\d(q).  
\ee
where $\om\cdot D=\om^i D_i$.

Using this result one can easily integrate over $q$ in 
(\ref{1700}) to get the 
heat kernel diagonal. After changing the integration 
variables $\om \to \sqrt t \om$ it takes the form \cite{avr94a,avr96a}
\bea
[U(t)]&=&(4\pi t)^{-d/2}\exp\left[-t\left(Q-{1\over 8}R
-{1\over 6} R_H\right)\right]\nonumber\\[12pt]
&&\times
(4\pi)^{-p/2}\int\limits_{\RR^p}d \om\, \b^{1/2}
\exp\left(-{1\over 4} <\om,\beta \om>\right)
\nonumber\\[12pt]
&&\times
\det^{1/2}_{{\rm Ad}({\cal H})}\left({\sinh(\sqrt t \om\cdot F/2)
\over \sqrt t \om\cdot F/2}\right)
\det^{-1/2}_{\End(TM)}\left({\sinh(\sqrt t \om\cdot D/2)
\over \sqrt t \om\cdot D/2}\right),
\label{2000}
\eea
where $\om\cdot F=\om^i F_i$, $F_i=\{F^j_{\ ik}\}$ are 
the generators of the holonomy algebra ${\cal H}$
in adjoint representation and
\be
R_H=-{1\over 4}\b^{ik}F^m_{\ il}F^l_{\ km}
\ee
is the scalar curvature of the holonomy group.

The remaining integration over $\om$ in (\ref{2000}) can be 
done in a rather {\it formal} way \cite{avr94b,avr95d}.
Let $a^{*}_i$ and $a_{k}$ be operators acting on a Hilbert space,
that form the following Lie algebra
\be
[a^j, a^{*}_k]=\delta^j_k,
\ee
\be
[a^i,a^k]=[a^{*}_i,a^{*}_k]=0.
\ee
Let $|0>$ be the `vacuum vector' in the Hilbert space, i.e.
\be
<0|0>=1,
\ee
\be
a^i|0>=0, \qquad <0|a^{*}_k=0.
\ee
Then the heat kernel (\ref{2000}) can be presented in an formal algebraic form 
without any integration
\bea
[U(t)]&=&(4\pi t)^{-d/2}
\exp\left[-t\left(Q-{1\over 8}R
-{1\over 6}R_H\right)\right]
\nonumber\\[12pt]
&&
\times\Big<0\Big|
\det^{1/2}_{{\rm Ad}({\cal H})}\left({\sinh(\sqrt t a\cdot F/2)\over 
\sqrt t a\cdot F/2}\right)
\det^{-1/2}_{\End(TM)}\left({\sinh(\sqrt t a\cdot D/2)\over 
\sqrt t a\cdot D/2}\right)
\nonumber\\[12pt]
&&
\times
\exp\left(<a^{*},\beta^{-1}a^{*}>\right)\Big|0\Big>. 
\eea
where $a\cdot F=a^k F_k$ and 
$a\cdot D=a^k D_k$.
This formal solution should be understood as a power 
series in the operators $a^k$ 
and $a^{*}_k$ and determines a well defined asymptotic 
expansion in $t\to 0$.

Let us stress that these formulas are 
{\it manifestly covariant} because they are expressed 
in terms of the invariants of the holonomy group $H$, 
i.e. the invariants of the Riemann curvature tensor. 
They can be used now to generate {\it all} HMDS-coefficients 
$[b_k]$ for {\it any} locally symmetric space, i.e. for any manifold 
with covariantly constant curvature, simply by expanding 
it in an asymptotic power series as $t\to 0$. 
Thereby one finds {\it all} covariantly constant terms in 
{\it all} HMDS-coefficients in a manifestly covariant way. 
This gives a very nontrivial example how the heat kernel 
can be constructed using only the Lie algebra of isometries 
of the symmetric space.

\section{Conclusion}

In present paper we have presented recent results  
in studying the heat kernel obtained in our papers 
\cite{avr86b,avr91b,avr93b,avr95b,avr94a,avr96a}.
We discussed some ideas connected with the problem of 
developing consistent covariant approximation schemes for 
calculating the heat kernel.
Especial attention is payed to the low-energy approxiamtion.
It is shown that  in the local analysis there exists
an algebraic structure (the Lie algebra of background 
jets) that turns out to be extremely useful
for the study of the low-energy approximation. 
Based on the background jets algebra
we have proposed a new promising approach for  calculating 
the low-energy heat kernel.

Within this framework we have obtained closed formulas 
for the heat kernel diagonal in the case of covariantly constant background. 
Besides, we were able to take into account the first 
and second derivatives of the endomorphism $Q$ in flat space.
The obtained formulas are manifestly covariant and 
applicable for a {\it generic} covariantly
constant background.
This enables to treat the results 
as the {\it generating functions} for the whole set of the 
Hadamard-\-Minakshisundaram-\-De~Witt-\-Seeley-\-coefficients. 
In other words, we have calculated {\it all} covariantly 
constant terms in {\it all} HMDS-\-coefficients.

Needless to say that the investigation of the low-energy 
effective action is of great importance in quantum gravity 
and gauge theories because it describes the dynamics of 
the vacuum state of the theory. 

\section*{Acknowledgments }

I would like to thank Professor Bor-Luh-Lin for his kind invitation
to present this talk and Professor Thomas Branson for
the hospitality expressed to me at the University of Iowa.
This work was supported by the Deutsche Forschungsgemeinschaft.


\begin{thebibliography}{99}

\bibitem{avr86b} {\sc I. G. Avramidi}, 
{\it Covariant Methods for the Calculation of 
the Effective Action in Quantum Field Theory and Investigation of 
Higher-Derivative Quantum Gravity}, 
PhD thesis, Moscow State University (1986), UDK 530.12:531.51, 178 pp. 
[in Russian]; Transl.: hep-th/9510140, 159 pp.

\bibitem{avr91b} {\sc I. G. Avramidi}, 
{\it A covariant technique for the calculation of the one-loop effective action}, 
Nucl. Phys. B {\bf 355} (1991) 712--754

\bibitem{avr93b} {\sc I. G. Avramidi}, 
{\it A new algebraic approach for calculating the heat kernel in gauge theories}, 
Phys. Lett. B {\bf 305} (1993) 27--34

\bibitem{avr95b} {\sc I. G. Avramidi}, 
{\it Covariant algebraic method for calculation of the low-energy heat kernel}, 
J. Math. Phys. {\bf 36} (1995) 5055--5070

\bibitem{avr94a} {\sc I. G. Avramidi}, 
{\it The heat kernel on symmetric spaces via integrating over the group of isometries}, 
Phys. Lett. B {\bf 336} (1994) 171--177

\bibitem{avr96a} {\sc I. G. Avramidi}, 
{\it A new algebraic approach for calculating the heat kernel in quantum gravity}, 
J. Math. Phys. {\bf 37} (1996) 374--394

\bibitem{avr95c} {\sc I. G. Avramidi}, 
{\it  New algebraic methods for calculating 
the heat kernel and the effective action in quantum 
gravity and gauge theories},
in: {\it `Heat Kernel Techniques and Quantum Gravity'}, 
Ed. S. A. Fulling, 
{\it Discourses in Mathematics and Its Applications},  
(College Station, Texas: Department of Mathematics, Texas A\& M University,
1995), pp.~115--140

\bibitem{avr94b} {\sc  I. G. Avramidi}, {\it Nonperturbative methods 
for calculating the heat kernel},
University of Greifswald (January, 1996), hep-th/9602169, 18 pp,
Proc. Int. Workshop `Global Analysis, Differential Geometry 
and Lie Algebras', Thessaloniki, Greece, Dec. 15-17, 1994;
to appear in: Algebras, Groups and Geometries

\bibitem{avr95d} {\sc  I. G. Avramidi}, {\it Covariant approximation schemes 
for calculation of the heat kernel in quantum field theory}, 
University of Greifswald (September, 1995), hep-th/9509075, 19 pp., Proc. 
Int. Seminar ``Quantum Gravity'', Moscow, June 12--19, to appear

\bibitem{barvinsky85} {\sc A. O. Barvinsky and G. A. Vilkovisky}, {\it The generalized
Schwinger--De~Witt technique in gauge theories and quantum gravity}, Phys. Rep.
C {\bf 119} (1985),  No 1, pp. 1--74.

\bibitem{berline92} {\sc N. Berline, E. Getzler and M. Vergne},
{\it Heat Kernels and Dirac Operators}, Springer, Berlin, 1992

\bibitem{branson90b} {\sc T. Branson, P. B. Gilkey and B. \O rsted},
{\it Leading terms in the heat invariants},
Proc. Amer. Math. Soc. {\bf 109} (1990)  437.

\bibitem{tam95}  {\sc S. A. Fulling}, Editor, {\it Heat Kernel Techniques and Quantum Gravity},
 Discourses in Mathematics and Its Applications,
No.~4, Department of Mathematics, Texas A\&M University,
College Station, Texas, 1995

\bibitem{gilkey75b} {\sc P. B. Gilkey}, 
{\it The spectral geometry of Riemannian manifold},
J. Diff. Geom. {\bf 10} (1975) 601--618.

\bibitem{gilkey95}
{\sc P. B. Gilkey}, {\it Invariance Theory, the Heat Equation
and the Atiyah-Singer Index Theorem},  FL:
Chemical Rubber Company, Boca Raton, 1995

\bibitem{hurt83} N. E. Hurt, {\it Geometric Quantization in Action: Applications of Harmonic Analysis
in Quantum Statistical Mechanics and Quantum Field Theory},
Reidel, Dordrecht, Holland, 1983

\end{thebibliography}
\end{document}